\documentclass[12pt]{article}
\usepackage{amsthm,amssymb,amsmath}
\usepackage[british]{babel}

 1 
1

\newcommand{\C}{\ensuremath{\mathbb{C}}}

\newcommand{\bt}{{\boldsymbol{t}}}
\newcommand{\ba}{{\boldsymbol{a}}}
\newcommand{\bT}{{\boldsymbol{T}}}

\renewcommand{\d}{\operatorname{d}}
\newtheorem{teh}{Theorem}

\newtheorem{lem}{Lemma}

\newcommand{\be}{\begin{equation}}
\newcommand{\ee}{\end{equation}}
\begin{document}

\title{\sc $\overline{\partial}$-equations, integrable deformations of
quasiconformal mappings and Whitham hierarchy
\thanks{ B. Konopelchenko  is supported in part by COFIN
2000 "S intesi" and L. Martinez Alonso by CICYT proyecto
PB98--0821  }}
\author{B. Konopelchenko$^{1}$ and L. Mart\'{\i}nez Alonso$^{2}$
 \\
\emph{ $^1$Dipartimento di Fisica, Universita
di Lecce}\\ \emph{73100 Lecce, Italy} \\
\emph{$^2$Departamento de F\'{\i}sica Te\'{o}rica II, Universidad
Complutense }\\\emph{ E-28040 Madrid, Spain}
 }
\date{} \maketitle
\begin{abstract}

It is shown that the dispersionless scalar integrable hierarchies
and, in general, the universal Whitham hierarchy are nothing but
classes of integrable deformations of quasiconformal mappings on
the plane. Examples of deformations of quasiconformal mappings
associated with explicit solutions of the dispersionless KP
hierarchy are presented.

\end{abstract}

\vspace*{.5cm}

\begin{center}\begin{minipage}{12cm}

\end{minipage}
\end{center}
\newpage

\section{Introduction}

It is widely recognize now that the dispersionless or
quasiclassical integrable equations and hierarchies form an
important part of the theory of integrable systems (see e.g.
\cite{1}-\cite{7}). They are the main ingredients of various
approaches to different problems which arise in physics and
applied mathematics,  as in the theory of topological quantum
fields \cite{8,9},  hydrodynamics  \cite{1} or optical
communications modelling \cite{10}. Moreover, it was recently
shown in \cite{11,12} that the dispersionless hierarchies provide
an effective tool to study some classical problems of the theory
of  conformal maps.

A purpose of this letter is to demonstrate the existence of an
intimate relation between dispersionless hierarchies and
quasiconformal mappings on the complex plane. We show that the
dispersionless scalar integrable hierarchies and, in general, the
members of the  universal Whitham hierarchy are nothing but
infinite-dimensional rings of integrable deformations of
quasiconformal mappings on the plane. The nonlinear
$\bar{\partial}$-equation
\begin{equation}\label{1}
S_{\bar{z}}=W\Big(z,\bar{z},S_{z}\Big),
\end{equation}
is the basic element of our analysis. Here $S(z,\bar{z},\bt)$ is a
complex-valued function depending an infinite set $\bt$ of
parameters (times), $S_{\bar{z}}:=\frac{\partial S}{\partial
\bar{z}},\;S_{z}:=\frac{\partial S}{\partial z}$ and $W$ is an
appropriate function of $z,\bar{z}$ and $S_z$. Equation \eqref{1}
is well-known in the theory of quasiconformal mappings. Namely,
under some mild conditions its solutions define quasiconformal
mappings on the plane (see \cite{13}-\cite{16}).

On the other hand, as it was shown in \cite{17,18} equation
\eqref{1} is a quasiclassical limit of the standard nonlocal
$\bar{\partial}$-problem arising in the $\bar{\partial}$-dressing
method, which allows us to generate the dispersionless
hierarchies. These two faces of equation \eqref{1} together with
some available rigorous results about the Beltrami equation, lead
to a natural derivation of the universal Whitham hierachy from
equation \eqref{1} as well as an exact solution method for
dispersionless hierarchies. To illustrate these facts we present
examples of exact solutions of the dKP hierarchy or, equivalently,
of explicit integrable deformations for the quasiconformal
mappings determined by equation \eqref{1}.

\section{Quasiconformal mappings}

Quasiconformal mappings are a natural and very rich extension of
the concept of conformal mappings.  For the sake of convenience we
remind here some of their basic properties (see e.g. \cite{13,14},
\cite{19}-\cite{23}).

A sense-preserving homeomorphism $w=f(z,\bar{z})$ defined on a
domain $G$ of the complex plane $\C$ is called
\emph{quasiconformal} (qc) if
\[
\frac{|f_z|+|f_{\bar{z}}|}{|f_z|-|f_{\bar{z}}|},
\]
is bounded on $G$. Here  $f_z$ and $f_{\bar{z}}$ are assumed to be
locally square-integrable  generalized derivatives. Since the
Jacobian of $f$ is assumed to satisfy
$J=|f_{z}|^2-|f_{\bar{z}}|^2>0$ almost everywhere in $G$, the
so-called \emph{complex dilatation} of $f$
$\mu_f:=\frac{f_{\bar{z}}}{f_{z}}$ verifies $|\mu_f|<1$ almost
everywhere in $G$. Thus, a natural analytic characterization of
qc-mappings can be given in terms of homeomorphic generalized
solutions of the linear Beltrami equation
\begin{equation}\label{2}
f_{\bar{z}}=\mu f_{z},
\end{equation}
where $\mu$ is any given measurable function $||\mu||_{\infty}<1$
on $G$. Obviously, for $\mu\equiv 0$ we get into the class of
conformal mappings. There is a qc-version of the Riemann's mapping
theorem for conformal mappings. Namely, for any
$||\mu||_{\infty}<1$ on a simply-connected domain $G$ there exists
a homeomorphism $w=f(z,\bar{z})$ which solves \eqref{2} and maps
$G$ onto a given simply-connected domain $\Delta$. A very simple
example of qc-mapping is the affine mapping $w=az+b\bar{z}+c$
where $a,b,c$ are complex constants with $|b|>|a|$. It maps every
circle into a ellipse.

The properties of solutions of the Beltrami equation \eqref{2} are
rather well studied (see e.g. \cite{19}). Some of them are
particularly important for our discussion. To present these
results we need to introduce the operators \cite{13}
\[
Th(z):=\frac{1}{2\pi\imath}\int\!\!\int_{\C\times\C}\frac{h(z')}{z'-z}\d
z' \wedge \d \bar{z}',\quad \Pi (z):=\frac{\partial Th}{\partial
z}(z),
\]
where the integral is taken in the sense of the Cauchy principal
value. Then one has \cite{13}:

\begin{lem}
For any $p>1$ the operator $\Pi$ defines a bounded operator in
$L^p(\C)$ and for any $0\leq k<1$ there exists $\delta>0$ such
that
\[
k||\Pi||_p<1,
\]
for all $|p-2|<\delta$.
\end{lem}

The next theorem summarizes the properties of qc-mappings that we
need in the subsequent discussion \cite{19,21,22}.

\begin{teh} Given a measurable function $\mu$ with compact support
inside the circle $|z|<R$ and such that $||\mu||_{\infty}<k<1$.
Then, for any fixed exponent $p=p(k)>2$ such that $k||\Pi||_p<1$,
it follows that

\begin{description}
\item[1)] There is a function $f_0$ on $\C$ with distributional
derivatives satisfying the Beltrami equation \eqref{2} such that
\begin{equation}\label{z}
f_0(z)=z+O(\frac{1}{z}),\quad z\rightarrow\infty,
\end{equation}
with $f_{0,{\bar z}}$ and $f_{0,z}-1$ being elements of $L^p(\C)$.
Any such $f_0$ is unique and determines a homeomorphism on $\C$
(the basic homeomorphism of \eqref{2}).
\item[2)] Every solution
of \eqref{2} on a domain $G$ of  $\C$ can be represented as
\begin{equation}\label{phi}
f(z)=\Phi(f_0(z)),
\end{equation}
where $\Phi$ is an arbitrary analytic function on the image
domain $f_0(G)$ of $G$ under the basic homeomorphism $f_0$.
\end{description}
\end{teh}

Deformations of qc-mappings have been discussed by several
authors. One can show \cite{13,20,22,23} that if
$\mu(z,\bar{z},\bt)$ depend analytically on one or several
parameters $\bt$, for fixed $z\in \C$, then the corresponding
family of basic homeomorphisms $f_0(z,\bar{z},\bt)$also depend
analytically on $\bt$.  However, it seems that apart from concrete
examples, as the isotropy deformation \cite{13,20} $\mu\rightarrow
t\mu$, most of this field remains to be investigated.

\section{Integrable deformations of qc-mappings and the universal
Whitham hierarchy}

Let us consider the $\bar{\partial}$-equation
\begin{equation}\label{5}
S_{\bar{z}}=W\Big(z,\bar{z},S_{z}\Big),
\end{equation}
where the function $W$ has a compact support in $\C$. The
infinitesimal symmetries $S\rightarrow S+ \epsilon f$ of \eqref{5}
satisfy the Beltrami equation
\begin{equation}\label{6}
f_{\bar{z}}=W'\Big(z,\bar{z},S_z\Big)f_z,
\end{equation}
where $W'(z,\bar{z},\xi):=W_{\xi}(z,\bar{z},\xi)$. By assuming
that the hypothesis of Theorem 1 are fulfilled, there is a unique
solution  $p$ of \eqref{6} on $\C$ (the basic homemorphism) such
that
\begin{equation}\label{7}
p=z+\frac{a_1}{z}+\frac{a_2}{z}+\ldots,\quad z\rightarrow\infty,
\end{equation}
and all other solutions of \eqref{6} are of the form
\begin{equation}\label{8}
f=\Phi(p),
\end{equation}
with $\Phi$ being an analytic function . This means that the set
of infinitesimal symmetries of \eqref{5} form an
infinite-dimensional ring.

Suppose we take a finite set of points $\{z_i\}_{i=1}^N$ outside
the support of the function $W$ in the complex plane, with
$z_1=\infty$ . We may consider solutions of \eqref{5}
$S(z,\bar{z},\bt)$ depending on an infinite  set of time
parameters $\bt:=(\bt_1,\ldots,\bt_N),\;
\bt_n:=(t_{1,1},t_{1,2},\ldots)$ , such that near the points
$\{z_i\}_{i=1}^N$ have a prescribed behaviour
\begin{equation}\label{ex}
S(z,\bar{z},\bt)=S_i(z,\bt_i)+\mbox{holomorphic part},\quad
z\rightarrow z_i,
\end{equation}
where the functions $S_i$ are singular at $z_i$.  Thus for every
time $t_A,\;( A=(i,n))$ we have a symmetry of \eqref{5}
\begin{equation}\label{9}
f_A=\frac{\partial S}{\partial t_A},
\end{equation}
and from Theorem 1 it follows that for any appropriate solution
$S(z,\bar{z},\bt)$  of \eqref{5} there exist a unique family of
basic homeomorphisms on $\C$
\[
p=z+\frac{a_1(\bt)}{z}+\frac{a_2(\bt)}{z}+\ldots,\quad
z\rightarrow\infty.
\]
We assume that $p$ can be characterized as $p(z,\bar{z},\bt):
=\frac{\partial S}{\partial t_{A_0}}$ for a certain time
parameter $t_{A_0}$. In this way, for any time $t_A$ there exists
a function $\Omega_A(p,\bt)$ such that
\begin{equation}\label{10}
\frac{\partial S}{\partial t_A}=\Omega_A(p,\bt).
\end{equation}
If $A=(i,n)$ then $\frac{\partial S}{\partial t_A}$ is a solution
of \eqref{6} on $G=\C^*-\{z_i\}$, so that $\Omega_A(p,\bt)$ is an
analytic function of $p$ on the image of $G$ under
$p(z,\bar{z},\bt)$.

The system \eqref{10} is an infinite set of Hamilton-Jacobi type
equations which characterizes  the infinite family of
deformations $p(z,\bar{z},\bt)$ of qc-mappings. By construction,
all the equations \eqref{10} are compatible, so that we have
\begin{equation}\label{12}
\frac{\partial \Omega_A}{\partial t_B}-\frac{\partial
\Omega_B}{\partial t_A}+\{\Omega_A, \Omega_B \}=0,
\end{equation}
where the Poisson bracket $\{,\}$ is given by
\begin{equation}\label{13}
\{F, G \}=\frac{\partial F}{\partial p}\frac{\partial G}{\partial
x}-\frac{\partial F}{\partial x}\frac{\partial G}{\partial p}.
\end{equation}
Equations \eqref{12} constitute examples of the so-called
universal Whitham hierarchy introduced in \cite{4}. Particular
choices of the functions $S_i$ in \eqref{ex} give rise to
different dispersionless hierarchies.

For instance, if we take a unique reference point $z_1=\infty$ and
assume
\begin{equation}\label{ass}
S(z,\bt)=\sum_{n\geq 1}z^n t_n+O(\frac{1}{z}),\quad
z\rightarrow\infty,
\end{equation}
we get the dKP hierarchy. In this case
\begin{equation}\label{pp}
p=\frac{\partial S}{\partial t_1}=z+O(\frac{1}{z}),\quad
z\rightarrow\infty.
\end{equation}
Moreover, as the functions $\frac{\partial S}{\partial t_i}$ are
solutions of \eqref{6} on $\C$,  their corresponding
representations $\Omega_i(p,\bt)$ are entire functions of $p$.
Therefore, according to the asymptotic behaviours \eqref{ass} and
\eqref{pp} it is clear that $\Omega_i(p,\bt)=(z^i)_+$, where
$z:=z(p,\bt)$ denotes the Laurent series  obtained by eliminating
$z$ as a function of $p$ in \eqref{pp}, and $(z^i)_+$ is the part
corresponding to the nonnegative powers of $p$ in the expansion of
$z^i$.. Hence, we conclude that
\begin{equation}\label{om}
\frac{\partial S}{\partial t_i}=(z^i)_+,\quad i\geq 2.
\end{equation}
The first two equations read
\begin{equation}\label{16}
\begin{gathered}
\frac{\partial S}{\partial t_2}=\Big(\frac{\partial S}{\partial
t_1}\Big)^2+u(\bt),\\\\
\frac{\partial S}{\partial t_3}=\Big(\frac{\partial S}{\partial
t_1}\Big)^3+\frac{3}{2}u(\bt)\,\frac{\partial S}{\partial
t_1}+\frac{3}{4}\,v(\bt),
\end{gathered}
\end{equation}
where $v_{t_1}=u_{t_2}$ and the function $u$ obeys the dKP
equation
\begin{equation}\label{17}
\Big(u_{t_3}-\frac{3}{2}u\,u_{t_1}\Big)_{t_1}=\frac{3}{4}u_{t_2t_2}.
\end{equation}
It must be noticed that by eliminating $u$ and $v$ in \eqref{16}
we get an autonomous partial differential equation for $S$
\begin{equation}\label{18}
\frac{\partial^2 S}{\partial t_1\partial
t_3}=\frac{3}{4}\,\frac{\partial^2 S}{\partial
t_2^2}+\frac{3}{2}\,\Big[\frac{\partial S}{\partial
t_2}-\Big(\frac{\partial S}{\partial
t_1}\Big)^2\Big]\frac{\partial^2 S}{\partial t_1^2}.
\end{equation}
It is just the first member of a hierarchy of autonomous equations
for $S$ which can be derived from \eqref{om}.

\section{Examples of qc-mappings and their dKP deformations }

To illustrate our analysis  let us consider the dKP hierarchy and
a $\bar{\partial}$-problem of the form
\begin{equation}\label{dbar}
 S_{\bar{z}}=\theta(1-z\bar{z})W_0\Big(S_z\Big),
\end{equation}
where $\theta(\xi)$ is the usual Heaviside function and $W_0(\xi)$
is an arbitrary differentiable function.  Observe that
\eqref{dbar} implies
\[
 m_{\bar{z}}=W'_0(m)m_z,\quad m:= S_z, \quad |z|<1,
\]
where $W'_0=\frac{\d W_0}{\d m}$. This equation can be solved at
once by applying the methods of characteristics, so that the
general solution $S_{in}$ of \eqref{dbar} inside the unit circle
$|z|<1$ is implicitly characterized by
\begin{equation}\label{gen}
\begin{gathered}
S_{in}=W_0(m)\bar{z}+mz-f(m),\\ \\W'_0(m)\bar{z}+z=f'(m),
\end{gathered}
\end{equation}
where $f=f(m)$ is an arbitrary differentiable function. Notice
that according to the second equation in \eqref{gen}, we have
\[
f'(m_0)=z,\quad m_0:=m(z,\bar{z})|_{\bar{z}=0},
\]
so that $f'(m_0)$ is the inverse function of $m_0=m_0(z)$.

The solution $S_{out}$ of \eqref{dbar} outside the unit circle is
any arbitrary analytic function ($\bar{z}$-independent). However,
in order to obtain a generalized solution of \eqref{dbar} with
locally square-integrable  generalized derivatives  we impose the
continuity of $S$ at the unit circle
\begin{equation}\label{out}
S_{out}(z)=S_{in}(z,\frac{1}{z}),\quad |z|=1.
\end{equation}
Moreover,  as we are dealing with the dKP hierarchy, we require
$S_{out}$ to be of the form
\begin{equation}\label{ss1}
S_{out}= \sum_{n\geq 1}z^n T_n+\sum_{n\geq 1}\frac{s_n(\bT)}{z^n}.
\end{equation}

In summary we may proceed as follows. Firstly we take a function
$m_0=m_0(z,\ba)$ depending on $z$ and a certain set of
undetermined parameters $\ba=(a_1,\ldots,a_n)$, then we get
$f=f(m,\ba)$ and solve for $m=m(z,\bar{z},\ba)$ in the second
equation of \eqref{gen}. The functions $S_{in}(z,\bar{z},\ba)$ and
$S_{out}(z,\ba)$ are determined by means of the first equation of
\eqref{gen} and \eqref{out}, respectively. Finally, we impose
$S_{out}$ to admit an asymptotic expansion of the form \eqref{ss1}
and find the parameters $\ba$ as functions of the dKP times $\bt$.
According to the results described in Section 3, if we get
solutions $S$ of \eqref{dbar} with enough regular generalized
derivatives $S_z$ and $S_{\bar z}$,  they will lead to solutions
of the dKP hierarchy which, for suitable values of the parameters
$\bt$, will determine deformations of qc-mappings. We observe that
the crucial property ensuring that a function $S$ of the form
\eqref{ss1} leads to a solution of the dKP hierarchy is that all
the derivatives $\frac{\partial S_{in}}{\partial t_i}$ can be
written as polynomials in $\frac{\partial S_{in}}{\partial t_1}$.
In this sense  it is helpful to notice that in our method $S$
depends on $\bt$ through the functions $\ba=\ba(\bt)$ so that from
\eqref{gen}
\[
\frac{\partial S_{in}}{\partial t_i}=- \frac{\partial
f(m,\ba(\bt))}{\partial t_i}.
\]
Therefore, a way to verify that $S$ leads to a dKP solution is to
check that  the derivatives $\frac{\partial
f(m,\ba(\bt))}{\partial t_i}$ can be expressed as polynomials in
$\frac{\partial f(m,\ba(\bt))}{\partial t_1}$.

  Let us
consider a couple of examples with $W_0(m)=m^2$. In this case
Equations \eqref{dbar} become
\begin{equation}\label{esc}
S_{in}=m^2\bar{z}+mz-f(m),\quad 2m\bar{z}+z=f'(m).
\end{equation}
\noindent
{\bf Example 1.} If we take $m_0= \frac{1}{a}(z-b)$
,then $f(m)=\frac{a}{2}m^2+bm+c$ and $m=\frac{z-b}{a-2\bar{z}}$.
Thus we get
\[
S=\left\{ \begin{array}{l}
\frac{1}{2}\frac{(z-b)^2}{a-2\bar{z}}-c,\quad |z|\leq 1\\ \\
\frac{1}{2}\frac{z(z-b)^2}{az-2}-c,\quad |z|\geq 1.
\end{array}
\right.
\]
Notice that the regularity of $S$ inside the unit circle requires
\begin{equation}\label{boun1}
|a|>2,
\end{equation}
which is in agreement with the required analyticity of $S$ on
$|z|>1$.

Now we have
\[
S=\frac{1}{2a}z^2+(\frac{1}{a^2}-\frac{b}{a})z+
\frac{2}{a^3}+\frac{b^2}{2a}-\frac{2b}{a^2}-c+O(\frac{1}{z}),\quad
z\rightarrow \infty,
\]
so that in order to fit with the dKP hierarchy  we have to
identify
\[
a=\frac{1}{2t_2},\quad b=2t_2-\frac{t_1}{2t_2},\quad
c=\frac{2}{a^3}+\frac{b^2}{2a}-\frac{2b}{a^2}.
\]
Notice that
\[
\frac{\partial f}{\partial t_i}=\frac{\partial a}{\partial
t_i}m^2+\frac{\partial b}{\partial t_i}m+\frac{\partial
c}{\partial t_i}.
\]
Hence, as $a$ is $t_1$-independent we have that $\frac{\partial
f}{\partial t_2}$ can be written as a quadratic polynomial in
$\frac{\partial f}{\partial t_1}$.
\newline

 \noindent {\bf Example 2.} If we take $m_0=az^2+bz+c$, then from
\eqref{esc} we get
\begin{equation}\label{f}
\begin{gathered}
f(m)=-\frac{b}{2a}m+\frac{1}{12
a^2}(4am+b^2-4ac)^{\frac{3}{2}}+d\\
=-\frac{b}{2a}m+\frac{1}{12 a^2}(4a\bar{z}m+2az+b)^3+d,
\end{gathered}
\end{equation}
and
\begin{equation}\label{m}
m=\frac{1}{8\bar{z}^2}\Big(\frac{1}{a}-4(z+\frac{b}{2a})\bar{z}-\sqrt{\frac{4}{a}
(\frac{b^2}{a}-4c)\bar{z}^2-\frac{8}{a}
(z+\frac{b}{2a})\bar{z}+\frac{1}{a^2}}\;\;\Big).
\end{equation}
Hence we have
\begin{equation}\label{s2}
S_{in}=(z+\frac{b}{2a})m+\bar{z}m^2-\frac{1}{12
a^2}(4a\bar{z}m+2az+b)^3-d.
\end{equation}
It is clear that in order to ensure the regularity of $S_{in}$ on
$|z|\geq 1$ and the analyticity of $S_{out}$ on $|z|>1$ we have to
require
\[
\begin{gathered}
\frac{4}{a} (\frac{b^2}{a}-4c)\bar{z}^2-\frac{8}{a}
(z+\frac{b}{2a})\bar{z}+\frac{1}{a^2}\neq 0,\quad |z|<1,\\\\
\frac{4}{a} (\frac{b^2}{a}-4c)\frac{1}{z^2}-\frac{8}{a}
(z+\frac{b}{2a})\frac{1}{z}+\frac{1}{a^2}\neq 0,\quad |z|\geq 1.
\end{gathered}
\]
One can see that this can be achieved provided
\[
\frac{|B|^2}{64}>(|A|+|B|+|C|),\quad
(\frac{|B|^2}{64}-|B|-|C|)^2>4|A|(\frac{|B|^2}{64}-|B|),
\]
where
\[
A=\frac{4}{a} (\frac{b^2}{a}-4c),\quad B=\frac{8}{a},\quad
C=-4\frac{b}{a^2}.
\]
Notice also that $S_{in}$ is regular at the origin as
\[
\lim_{z\rightarrow 0}m=\frac{1}{2}-8\frac{C^2}{B^3}.
\]

In order to introduce the dKP times in our solution it is helpful
to use the following identity valid on the unit circle
$\bar{z}=z^{-1}$
\begin{equation}\label{iddd}
S_{out,z}=
S_{in,z}-\frac{1}{z^2}S_{in,\bar{z}}=m-\Big(\frac{m}{z}\Big)^2.
\end{equation}
Thus from the expansion
\[
\begin{gathered}
m(z,z^{-1})=-\Big(\frac{1}{2}+\frac{1}{8a}(1-\sqrt{1-8a})\Big)
z^2+\\\\
\frac{b}{4a}\Big(-1+\frac{1}{\sqrt{1-8a}}\Big)z+
\frac{-2b^2+(-1+8a)c}{(-1+8a)\sqrt{1-8a}}+\ldots,\quad
z\rightarrow\infty,
\end{gathered}
\]
and by setting $S_{out,z}=3t_3z^2+2t_2z+t_1+\ldots$ in
\eqref{iddd} we get
\[
\begin{gathered}
a=\frac{1+36\,t_3-{\sqrt{1-36\,t_3+432\,t_3^2-1728\,t_3^3}}}{2\,
\left(9+72\,t_3+144\,t_3^2\right)},\\\\
b=\frac{-32\,{\sqrt{1-8\,a}}\,a^2t_2}{\left(-1+{\sqrt{1-8\,a}}\right)\,
\left(-1+{\sqrt{1-8\,a}}+8\,a\right)},\\\\
c=32\,{\sqrt{1-8\,a}}\,
[\left(a-8\,a^2\right)\,\left(1-{\sqrt{1-8\,a}}+
4\,\left(-2+{\sqrt{1-8\,a}}\right)\,a+8a^2\right)\,t_1+\\
16\,a^3\left(1-{\sqrt{1-8\,a}}+4\left(-2+{\sqrt{1-8\,a}}\right)
\,a\right)\,t_2^2]\times
\\
[\left(-1+{\sqrt{1-8\,a}}\right)^2\left(-1+{\sqrt{1-8\,a}+8\,a}\right)^3]^{-1}
\end{gathered}
\]
Observe that $a$ and $b$ are $t_1$-independent so that
\[
\frac{\partial f}{\partial
t_1}=-\frac{1}{2a}\sqrt{4am+b^2-4ac}\,\frac{\partial c}{\partial
t_1}+\frac{\partial d}{\partial t_1}.
\]
Hence it follows at once that the derivatives of $f$ with respect
to $t_2$ and $t_3$ can be written as polynomials in
$\frac{\partial f}{\partial t_1}$.

\hspace{1cm}

\noindent{\bf Aknowledgements}

We would like to thank Prof. E. Medina for carrying out the
computer calculation of the coefficients in the second example of
Section 4.

\end{document}